\input harvmac
\input epsf
%
%
\newcount\figno
\figno=0
\def\fig#1#2#3{
\par\begingroup\parindent=0pt\leftskip=1cm\rightskip=1cm\parindent=0pt
\baselineskip=11pt \global\advance\figno by 1 \midinsert
\epsfxsize=#3 \centerline{\epsfbox{#2}} \vskip 12pt {\bf Fig.\
\the\figno: } #1\par
\endinsert\endgroup\par
}
\def\figlabel#1{\xdef#1{\the\figno}}
\overfullrule=0pt
\parskip=0pt plus 1pt
\sequentialequations

\def\half{{1\over 2}}

\newcount\figno
\figno=0
\def\fig#1#2#3{
\par\begingroup\parindent=0pt\leftskip=1cm\rightskip=1cm\parindent=0pt
\baselineskip=11pt
\global\advance\figno by 1
\midinsert
\epsfxsize=#3
\centerline{\epsfbox{#2}}
\vskip 12pt
\centerline{Fig.\ \the\figno: #1}\par
\endinsert\endgroup\par
}
\def\figlabel#1{\xdef#1{\the\figno}}
\def\np{Nucl.\ Phys.}
\def\pl{Phys.\ Lett.}

\def\cmp{Comm.\ Math.\ Phys.}

\def\jhep{JHEP} 
\Title{\vbox{\baselineskip12pt\hbox{hep-th/0406259}%
\hbox{HRI-P-0406003}%
\hbox{UT-04-19}
}}%
{\vbox{\centerline{Exact Noncommutative Solitons}
\vskip8pt
\centerline{in $p\,$-Adic Strings and BSFT}}}

{\vskip -20pt\baselineskip 14pt

\centerline{
Debashis Ghoshal\foot{On sabbatical leave from the Harish-Chandra 
Research Institute, Chhatnag Road, Allahabad 211 019.}} 

\bigskip
\centerline{\it Department of Physics, University of Tokyo}
\centerline{\it 7-3-1 Hongo, Bunkyo-ku, Tokyo 113 0033, Japan}
\smallskip
\centerline{\tt ghoshal@hep-th.phys.s.u-tokyo.ac.jp}
\smallskip

\vglue .3cm
\bigskip\bigskip

\noindent The tachyon field of $p$-adic string theory is made 
noncommutative by replacing ordinary products with noncommutative
products in its exact effective action. The same is done for the
boundary string field theory, treated as the $p\to 1$ limit of the
$p$-adic string. Solitonic lumps corresponding to D-branes are
obtained for all values of the noncommutative parameter
$\theta$. This is in contrast to usual scalar field theories in which
the noncommutative solitons do not persist below a critical value of
$\theta$. As $\theta$ varies from zero to infinity, the solution
interpolates smoothly between the soliton of the $p$-adic theory
(respectively BSFT) to the noncommutative soliton.

}
\Date{June 2004}

\lref\SenRev{A.\ Sen, {\it Non-BPS states and branes in 
string theory}, [{\tt hep-th/9904207}].}

\lref\NCTach{K.\ Dasgupta, S.\ Mukhi and G.\ Rajesh, {\it 
Noncommutative tachyons}, \jhep\ {\bf 0006} (2000) 022 
[{\tt hep-th/0005006}];
\hfill\break
J\ Harvey, P.\ Kraus, F.\ Larsen and E.\ Martinec, {\it
D-branes and strings as non-commutative solitons}, \jhep\
{\bf 0007} (2000) [{\tt hep-th/0005031}]\semi
E.~Witten,
{\it Noncommutative tachyons and string field theory},
arXiv:hep-th/0006071.
} 

\lref\FrOl{P.G.O.\ Freund and M.\ Olson,
{\it Nonarchimedean strings}, \pl\ {\bf B199} (1987) 186.}

\lref\FrWi{P.G.O.\ Freund and E.\ Witten, {\it Adelic string 
amplitudes}, \pl\ {\bf B199} (1987) 191.}

\lref\BFOW{L.\ Brekke, P.G.O.\ Freund, M.\ Olson and 
E.\ Witten, {\it Non-archimedean string dynamics},
Nucl.\ Phys.\  {\bf B302} (1988) 365.}

\lref\OtherPadic{P.H.\ Frampton and Y.\ Okada,
{\it The p-adic string N point function},
Phys.\ Rev.\ Lett.\  {\bf 60} (1988) 484;
\hfill\break
{\it Effective scalar field theory of p-adic string},
Phys.\ Rev.\  {\bf D37} (1988) 3077; \hfill\break
P.H.\ Frampton, Y.\ Okada and M.R.\ Ubriaco,
{\it On adelic formulas for the p-adic string},
Phys.\ Lett.\  {\bf B213} (1988) 260.}

\lref\Nonlocal{
B.L.\ Spokoiny,
{\it Quantum geometry of non-archimedean particles and strings},
Phys.\ Lett.\  {\bf B208} (1988) 401; \hfill\break
G.\ Parisi,
{\it On p-adic functional integrals},
Mod.\ Phys.\ Lett.\  {\bf A3} (1988) 639;\hfill\break
R.B.\ Zhang,
{\it Lagrangian formulation of open and closed p-adic strings},
Phys.\ Lett.\  {\bf B209} (1988) 229.}

\lref\Zabro{
A.V.\ Zabrodin,
{\it Non-archimedean strings and Bruhat-Tits trees},
Commun.\ Math.\ Phys.\  {\bf 123} (1989) 463; \hfill\break
L.O.\ Chekhov, A.D.\ Mironov and A.V.\ Zabrodin, {\it
Multiloop calculations in p-adic string theory and Bruhat-Tits trees},
Commun.\ Math.\ Phys.\  {\bf 125} (1989) 675.}

\lref\FrNi{
P.H.\ Frampton and H.\ Nishino,
{\it Stability analysis of p-adic string solitons},
Phys.\ Lett.\  {\bf B242} (1990) 354.}

\lref\MaZa{
A.V.\ Marshakov and A.V.\ Zabrodin,
{\it New p-adic string amplitudes},
Mod.\ Phys.\ Lett.\  {\bf A5} (1990) 265.}

\lref\Chekhov{
L.O.\ Chekhov and Y.M.\ Zinoviev, {\it
p-Adic string compactified on a torus},
Commun.\ Math.\ Phys.\ {\bf 130} (1990) 130.}

\lref\BrFrRev{
L.\ Brekke and P.G.O.\ Freund, {\it p-Adic numbers in physics}, 
Phys.\ Rep.\ {\bf 133} (1993) 1, and references therein.}

\lref\GhSeP{
D.\ Ghoshal and A.\ Sen, {\it Tachyon condensation and brane descent
relations in $p$-adic string theory}, \np\ {\bf B584} (2000) 
300, [{\tt hep-th/0003278}].}

\lref\MinaWH{
J.A.~Minahan,
{\it Mode interactions of the tachyon condensate in p-adic string theory},
JHEP {\bf 0103}, 028 (2001), [{\tt hep-th/0102071}].}

\lref\MinaPD{
J.A.~Minahan,
{\it Quantum corrections in p-adic string theory}, {\tt hep-th/0105312}.
}

\lref\MZVX{
N.~Moeller and B.~Zwiebach,
{\it Dynamics with infinitely many time derivatives and rolling tachyons},
JHEP {\bf 0210}, 034 (2002), [{\tt hep-th/0207107}].
}

\lref\YangNM{
H-t.~Yang,
{\it Stress tensors in p-adic string theory and truncated OSFT},
JHEP {\bf 0211}, 007 (2002), [{\tt hep-th/0209197}].
}

\lref\VolovichZH{
Y.~Volovich, {\it Numerical study of nonlinear equations with infinite
number of derivatives}, J.\ Phys.\ A {\bf 36}, 8685 (2003), [{\tt
math-ph/0301028}].
}

\lref\MoellerGG{
N.~Moeller and M.~Schnabl,
{\it Tachyon condensation in open-closed p-adic string theory},
{\tt hep-th/0304213}.
}

\lref\VlVo{
V.S.~Vladimirov and Y.I.~Volovich,
{\it On the nonlinear dynamical equation in the p-adic string theory},
{\tt math-ph/0306018}.
}

\lref\Volo{
I.V.\ Volovich,
{\it p-Adic string},
Class.\ Quant.\ Grav.\  {\bf 4} (1987) L83; \hfill\break
B.\ Grossman,
{\it p-Adic strings, the Weyl conjectures and anomalies},
Phys.\ Lett.\  {\bf B197} (1987) 101.}

\lref\GeSh{
A.A.~Gerasimov and S.L.~Shatashvili,
{\it On exact tachyon potential in open string field theory},
JHEP {\bf 0010}, 034 (2000),
[{\tt hep-th/0009103}].
}

\lref\WiBSFT{
E.\ Witten,
{\it On background independent open string field theory},
Phys.\ Rev.\  {\bf D46}, 5467 (1992)
[{\tt hep-th/9208027}];\hfill\break
E.\ Witten, {\it
Some computations in background independent off-shell string theory},
Phys.\ Rev.\  {\bf D47}, 3405 (1993)
[{\tt hep-th/9210065}].}

\lref\ShBSFT{
S.L.\ Shatashvili, {\it
Comment on the background independent open string theory},
Phys.\ Lett.\  {\bf B311}, 83 (1993)
[{\tt hep-th/9303143}];\hfill\break
S.L.\ Shatashvili, {\it
On the problems with background independence in string theory},
{\tt hep-th/9311177}. }

\lref\KuMaMo{
D.~Kutasov, M.~Marino and G.~Moore,
{\it Some exact results on tachyon condensation in string field theory},
JHEP {\bf 0010}, 045 (2000)
[{\tt hep-th/0009148}].
}

\lref\GhSeN{
D.~Ghoshal and A.~Sen, {\it Normalisation of the background
independent open string field theory action}, JHEP {\bf 0011}, 021
(2000) [{\tt hep-th/0009191}]\semi
D.~Ghoshal,
{\it Normalization of the boundary superstring field theory action},
in Strings 2001, Eds.\ A.\ Dabholkar {\it et al}, AMS (2002)
[{\tt hep-th/0106231}].
}

\lref\Cornalba{
L.~Cornalba,
{\it Tachyon condensation in large magnetic fields with background 
independent string field theory},
Phys.\ Lett.\ B {\bf 504}, 55 (2001)
[arXiv:hep-th/0010021].
}

\lref\Okuyama{
K.~Okuyama,
{\it Noncommutative tachyon from background independent open string 
field theory},
Phys.\ Lett.\ B {\bf 499}, 167 (2001)
[arXiv:hep-th/0010028].
}

\lref\DouglasBA{
M.R.~Douglas and N.A.~Nekrasov,
{\it Noncommutative field theory},
Rev.\ Mod.\ Phys.\  {\bf 73}, 977 (2001)
[{\tt hep-th/0106048}].
}

\lref\HarveyYN{
J.~A.~Harvey,
{\it Komaba lectures on noncommutative solitons and D-branes},
{\tt hep-th/0102076}.
}

\lref\GoMiSt{
R.\ Gopakumar, S.\ Minwalla and A.\ Strominger,
{\it Noncommutative solitons}, \jhep\ {\bf 0005} (2000) 020.
[{\tt hep-th/0003160}].}

\lref\Schom{
V.~Schomerus,
{\it D-branes and deformation quantization},
JHEP {\bf 9906}, 030 (1999)
[{\tt hep-th/9903205}].
}

\lref\GhKa{
D.~Ghoshal and T.~Kawano, work in progress.}

\lref\NCSoln{
A.P.~Polychronakos,
{\it Flux tube solutions in noncommutative gauge theories},
Phys.\ Lett.\ {\bf B495}, 407 (2000)
[{\tt hep-th/0007043}]\semi
D.~P.~Jatkar, G.~Mandal and S.~R.~Wadia,
{\it Nielsen-Olesen vortices in noncommutative Abelian Higgs model},
JHEP {\bf 0009}, 018 (2000)
[{\tt hep-th/0007078}]\semi
J.~Harvey, P.~Kraus and F.~Larsen,
{\it Exact noncommutative solitons},
JHEP {\bf 0012}, 024 (2000)
[{\tt hep-th/0010060}].
}

\lref\GoHeSp{
R.\ Gopakumar, M.\ Headrick and M.\ Spradlin, {\it On
noncommutative multisolitons}, \cmp\ {\bf 233} (2003) 355,
[{\tt hep-th/0103256}].}

\lref\DrVo{
B.\ Dragovich and I.\ Volovich, {\it $p$-Adic strings and
noncommutativity}, in Noncommutative structures in Mathematics \&\
Physics, S.\ Duplij and J.\ Wess (Eds.), Kluwer (2001), 391--399.}

\lref\moscow{
D.~Ghoshal, {\it Noncommutative p-Tachyon}, to appear in the
Proceedings of {\it p-Adic MathPhys 2003}, Moscow. }

{\nopagenumbers

\ftno=0
}

\newsec{Introduction}
The progress in our understanding of the tachyon field localized on an
unstable D-brane, led by the pioneering work of Sen (see
\refs{\SenRev} and references therein), has opened a new window into
the dynamics of string theory. However, the full string theory being
intractible, one hopes to gain insight into its qualitative features
by studying simpler models which share its essential properties. The
$p$-adic string theory, introduced in \refs{\FrOl} and studied further
in \refs{\FrWi\BFOW\OtherPadic\Nonlocal\Zabro\FrNi\MaZa--\Chekhov}
(see \refs{\BrFrRev} for a review), is one such model on which one has
considerable analytic control. Indeed, in \refs{\GhSeP} it was shown
that the tree level effective action of the $p$-adic tachyon, the
expression for which is known exactly, provides an explicit
realization of the Sen conjectures. This, and other properties, were
explored further in
\refs{\MinaWH\MinaPD\MZVX\YangNM\VolovichZH\MoellerGG--\VlVo}. 
Although the $p$-adic string itself is an exotic object, the spacetime
it describes is the familiar one\foot{For a different, perhaps more
exotic, type of $p$-adic string, see \refs{\Volo}}. Therefore, field
theoretical tools can be applied profitably.

Moreover, it turns out\refs{\GeSh} that a limit in which $p$ tends to
one, approximates the boundary string field theory (BSFT) description
of the usual bosonic string\refs{\WiBSFT,\ShBSFT}. More precisely,
upto two derivatives, the effective action of the tachyon can be
computed exactly in BSFT. This matches with the $p\to 1$ limit of the
tachyon of the $p$-adic string. Many of the features of the $p$-adic
theory, {\it e.g.}, the existence of gaussian lump solitons, survive
this limit. The results obtained in BSFT are in accordance with the
conjectures of Sen\refs{\KuMaMo,\GhSeN}.

In spite of these progress, $p$-adic string still remains an enigmatic
entity. We certainly lack an understanding of $p$-adic closed
string. Even open string modes, other than the tachyon, are very
poorly understood. It is therefore almost impossible to address
questions about $p$-adic string in non-trivial backgrounds, due either
to closed strings, like curved spacetime, or to open strings, like an
electromagnetic field. It should be recalled that thanks to
Refs.\Zabro, we do have a `worldsheet' action for the $p$-adic
string. However, the `worldsheet' being a tree, {\it i.e.}, an
infinite Bethe lattice, the interpretation of it is difficult.
Consequently the worldsheet approach, so fruitful for usual strings,
is yet to be exploited to our advantage.

We do, however, know of one background, namely a constant background
of the antisymmetric second rank tensor field $B$, that leads to a
simple modification of the spacetime effective action. One can use the
same form of the action, but use a noncommutative product, instead of
the ordinary product, in multiplying fields. There is a vast
literature on the subject, see Ref.\DouglasBA\ for a review and
references. In particular, it leads to a great simplification in
undertstanding tachyon condensation. Generic noncommutative scalar
field theories admit solitonic lump solutions\refs{\GoMiSt}, which can
be identified with D-branes\refs{\NCTach} (see also the review
\refs{\HarveyYN}).

In this paper, we propose to modify the spacetime action of the
$p$-adic tachyon by replacing the ordinary product of fields by the
noncommutative Moyal product. This will be our definition of the
noncommutative $p$-adic tachyon. We will study the resulting equations
of motion and find gaussian lump solutions for {\it all} values of the
noncommutative parameter $\theta$. These are shown to interpolate from
the usual $p$-adic solitons to the soliton found in
Ref.\refs{\GoMiSt} in the limit of infinitely large noncommutativity.
Next we consider the $p\to 1$ limit\GeSh, which is known to be the
bosonic string in the boundary string field theory (BSFT)
formalism\refs{\WiBSFT\ShBSFT\KuMaMo--\GhSeN}.  Exact noncommutative
soltions are obtained for a noncommutative
deformation\refs{\Cornalba,\Okuyama} of this theory. We will also
comment on multisolitons.

In the following, we sometimes use the economical, if somewhat
confusing, abbreviations $p$-tachyon, $p$-string and $p$-soliton for
the tachyon, string and soliton respectively of the $p$-adic string
theory. In the $p\to 1$ limit, these naturally connect to the usual
tachyon, etc!

The solitons of the noncommutative $p$-adic string theory were
presented in the conference {\it p-Adic MathPhys 2003} held at the
Steklov Institute, Moscow\moscow.


\newsec{Review of the $p$-adic tachyon} 
In $p$-adic string theory {\it all} tree level amplitudes involving
tachyons in the external states can be computed. This makes it
possible to write the exact effective action for the $p$-tachyon
field. It was computed in \refs{\BFOW} and is described by the
lagrangian
\eqn\paction{
{\cal L}_p\; =\;-\half\varphi\,p^{-\half\lform}\varphi + 
{1\over p+1}\,\varphi^{p+1}. }
Although this was arrived at by computing Koba-Nielsen
amplitudes in the $p$-adic theory, which is meaningful and possible 
only for a prime $p$, the final spacetime action makes sense for
all integer values of $p$. The potential has a local minimum
and two (respectively one) local maxima for odd (respectively
even) integer $p$. 
There are always runaway pathological singularities. 


The equation of motion following from the above is
\eqn\paeom{
p^{-\half\lform}\varphi = \varphi^p. }
Apart from the trivial constant solutions for $\varphi=0,1$, this 
admits soliton solutions. In fact, the equations separate in 
the arguments and for any (spatial) direction $y$, we get 
\eqn\plump{
\varphi(y) = p^{1/2(p-1)}\exp\left(-\,{p-1\over2p\ln p}\,
y^2\right), }
a gaussian lump whose amplitude and spread are correlated. 
While the solutions were already found by the authors of
Ref.\refs{\BFOW}, these were identified as the D-branes 
of the $p$-string theory in \refs{\GhSeP}.

More specifically, considering a lump along the last $25-m$
dimensions, one has a configuration in which energy is 
localized in codimension $25-m$. This is to be identified 
with the D-$m$-brane of the $p$-string theory. There
are branes for all values $m=0,1,\cdots,25$ satisfying descent
relations conjectured by Sen\refs{\SenRev}. In particular,
the ratios of their tensions match that of the energies 
of the corresponding solitons. It was also shown that one 
gets a consistent truncation of the worldvolume theory on 
a D-$m$-brane by keeping only the tachyon mode, the resulting
theory is exactly of the form of the parent theory in $(m+1)$
dimensions.   


\newsec{A noncommutative deformation of the $p$-adic tachyon}
In this section, we will study the $p$-adic string in a nontrivial
background, namely the one produced by a constant two-form
antisymmetric tensor field $B$. In usual string theory, this has the
effect that spatial coordinates no longer commute, instead
\eqn\noncomspace{
[x^i,x^j] = i\theta\epsilon^{ij},}
where, $\theta$ is related to the constant value of $B_{ij}$.
For simplicity, let us restrict our attention to the minimal case 
in which two spatial directions, say $x^1$ and $x^2$ fail to 
commute. An equivalent description of the Physics is through the
use of commuting coordinates, but while multiplying fields,
which are dependent on $x^1$ and $x^2$, one uses the Moyal 
star product
\eqn\moyalprod{
f(x^1,x^2)\star g(x^1,x^2) = f(x^1,x^2)\exp\left({i\over2}
\theta\epsilon^{ij}{\buildrel\leftarrow\over \partial_i}
{\buildrel\rightarrow\over\partial_j}\right) g(x^1,x^2), }
instead of the ordinary pointwise multiplication. Alternatively, $f,
g$ may be thought of as operator valued functions through the
well-known Moyal-Weyl correspondence\refs{\DouglasBA,\HarveyYN}.

Ordinary commutative field theories may be deformed by the use of the
Moyal star product. These are known to exhibit rather interesting
properties, both perturbative as well as nonperturbative. For example,
nontrivial soliton solutions were found\refs{\GoMiSt} in a
noncommutative scalar field theory with generic polynomial potential
in the limit of infinite noncommutativity: $\theta\to\infty$. These
solutions owe their existence to the infinite number of derivatives
that appear in the star product. They are essentially the solutions of
the equation
\eqn\projector{
\phi\star\phi\sim\phi
}
defining projectors. The simplest of the solitons is a gaussian 
lump whose width and amplitude are fixed:
\eqn\nclump{
\phi(x^1,x^2) = 2 \exp\left(-{(x^1)^2+(x^2)^2\over\theta}
\right).
}
These solitons become unstable at finite values 
of $\theta$. 

On the other hand, the effective action \paction\ of the $p$-tachyon
field already contains an infinite number of derivatives. Moreover,
there are gaussian lump solutions of the equations of motion.  This
leads to the hope that a noncommutative deformation of the $p$-tachyon
may admit solitons that are stable at all values of the
noncommutative parameter $\theta$. We will see that this indeed
turns out to be the case for the gaussian soliton.

In order to make a noncommutative deformation, we shall follow the
standard practice and replace ordinary products of the $p$-tachyon field
$\varphi$ by the star product \moyalprod\ in the action
\paction. This gives the action for the noncommutative $p$-tachyon:
\eqn\ncpaction{
{\cal L}_p^{NC}\; =\; -\,\half\,\varphi\,\star\,p^{-\half\lform}\varphi 
+ {1\over p+1}\,(\star\varphi)^{p+1}, }
where we have used an obvious shorthand notation for the $(p+1)$-fold
noncommutative product of $\varphi$. For us, the action \ncpaction\
{\it defines} the noncommutative $p$-tachyon. It will, however, be
desirable, using the results of \refs{\Zabro}, to develop a worldsheet
understanding of this deformation along the lines of
\refs{\Schom}. 

The equation of motion, as usual, is the noncommutative variant 
of \paeom:
\eqn\ncpeom{
p^{-\half\lform}\varphi = (\star\varphi)^p. }
We will only be interested in the part dependent on $x^1$ 
and $x^2$. It will, therefore, suffice to restrict our attention to 
these directions only. Hence $\varphi=\varphi(x^1,x^2)$ and 
$\lform=\partial_1^2 +\partial_2^2$ for us.

The trivial solutions, $\varphi=0,1$ describing constant configurations,
are obviously still solutions of \ncpeom.  More interestingly, there is
a nontrivial solution. This is a gaussian solitonic lump and may
be obtained by Fourier transformation on a gaussian ansatz. To this
end, let us note that the $n$-fold star product of the gaussian
\eqn\Gansatz{
g(\hbox{\bf x}) = A^2 \exp\left(-ar^2\right),
\qquad\hbox{\bf x}=(x^1,x^2),\;\;r^2=(x^1)^2+(x^2)^2, }
is again a gaussian albeit with modified width and amplitude:
\eqn\gaussstarn{
\left(\star g\right)^n(\hbox{\bf x}) = { A^{2n}\over
\displaystyle\sum_{i=0}^{\lfloor n/2\rfloor}\left(
{n\atop 2i}\right)\left(a\theta\right)^{2i}}\; \exp\left[ -\, {
\displaystyle\sum_{i=0}^{\lfloor(n-1)/2\rfloor}\left(
{n\atop 2i+1}\right)\left(a\theta\right)^{2i}\over
\displaystyle\sum_{i=0}^{\lfloor n/2\rfloor}\left(
{n\atop 2i}\right)\left(a\theta\right)^{2i} }\, 
ar^2\right]. }
It is easy to obtain the above by induction. The differential operator
on the LHS of \ncpeom\ also modifies the width and amplitude of the
gaussian \Gansatz. Thus one gets a solution to the equation of motion
by equating the width and amplitude of the gaussian on both sides of
\ncpeom. 

The width $a$ is determined by a polynomial of degree $p$:
\eqn\apolynom{
\sum_{i=0}^{\left\lfloor p/2\right\rfloor}\left({p\atop 2i}
\right)(a\theta)^{2i} - (1 - 2a\ln p)
\sum_{i=0}^{\left\lfloor (p-1)/2\right\rfloor}\left(
{p\atop 2i+1}\right)(a\theta)^{2i} = 0.} 
For an odd integer $p$, one is guaranteed to have one real root.  (For
$p=2$, the roots turn out to be real, the positive root being the
relevant one.)  While it is not possible to determine $a$ explicitly
as a function of $\theta$ for most $p$, we can check that the known
limits are recovered. In the commutative limit $\theta=0$, the
polynomial reduces to a linear equation in $a$, the solution of which
is the result $a={p-1\over 2p\ln p}$ in Eq.\plump.  On the other hand,
in the limit of infinite noncommutativity $\theta\to\infty$, one
cannot naively keep only the term of highest degree in $a$. Using the
fact that $a\sim 1/\theta$ in this case\GoMiSt, this term is actually
non-leading. It is easy to check that the constant of proportionality
is one, thus one gets the value in Eq.\nclump. The polynomial
\apolynom\ for $p=3$ is plotted in Fig.~2 for different values of
$\theta$. For increasing $\theta$, the root moves towards the origin,
as expected.

The amplitude likewise is determined in terms of $a$ from 
\eqn\amplitude{
A = {\left[\sum_{i=0}^{\left\lfloor(p-1)/2\right\rfloor}
\left({p\atop 2i+1}\right)
\left(a\theta\right)^{2i}\right]}^{1/2(p-1)},
} 
which, again, interpolates between the values $A=p^{1/2(p-1)}$ 
and $A=\sqrt{2}$ in the commutative and noncommutative limits
respectively. (In the case of $p=2$, the amplitude is a constant
$A=\sqrt{2}$ independent of $\theta$.)

\fig{Polynomial \apolynom\ for $p=3$ for different $\theta$ 
(increasing upwards).}{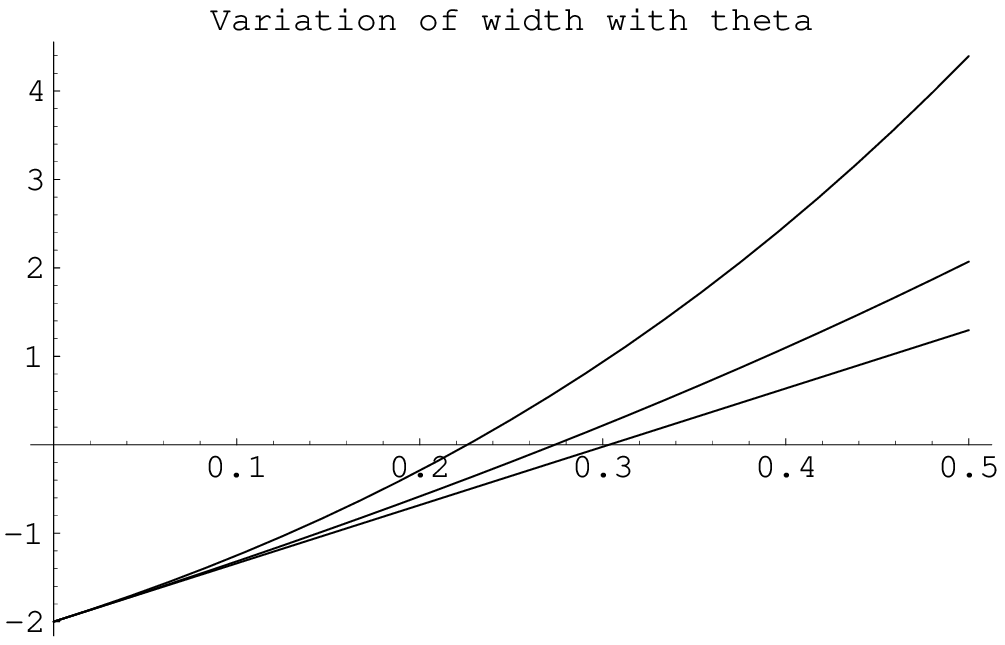}{3.5truein}

In summary, we get, for all values of $\theta$, a gaussian
lump solution which interpolates smoothly between the soliton
\plump\ of the $p$-string theory on one hand and the noncommutative
GMS soliton \nclump\ on the other hand.

It is easy to check that the total energy (see \refs{\MZVX,\YangNM}
for the definition of energy in $p$-string theory) of the
$p$-soliton
\eqn\energy{
E = {p-1\over2(p+1)}\int d^2x\, \left(\star g\right)^{p+1}
= {\pi(p-1)A^{2p+2}\over2(p+1)}\left[a\sum_{i=0}^{\lfloor p/2\rfloor}
\left({p+1\atop 2i+1}\right)\left(a\theta\right)^{2i}\right]^{-1}, }
with $g$ given by \Gansatz, \apolynom\ and \amplitude\ is a 
function of $\theta$, with $E\sim{\cal O}(1)$ for $\theta=0$ and 
$E\sim\theta$ for large values of the noncommutative parameter.


\newsec{The $p\to 1$ limit and BSFT}
Let us consider the $p\to 1$ limit now. In this limit, the 
lagrangian \paction\ of the $p$-string theory reduces 
to\GeSh\foot{Curiously, Spokoiny\Nonlocal\ considered this limit
and speculated on its relation to the usual string theory already 
in the early days of $p$-adic string theory!}
\eqn\bsftaction{
{\cal L}_{p\to 1} =
\half\varphi\,\lform\varphi + \half\varphi^2\left(\ln
\varphi^2 - 1\right).
} 
Curiously, this is exactly the action of the usual bosonic string, 
truncated to two derivatives, in the boundary string field theory
(BSFT)\refs{\WiBSFT,\ShBSFT}\ approach. 

A noncommutative deformation of this theory is described by the
lagrangian
\eqn\ncbsftaction{
{\cal L}_{p\to 1}^{NC} = 
\half\varphi\star\lform\,\varphi + \half\varphi\star\varphi\star
\left(\ln_\star(\varphi\star\varphi) - 1\right)
} 
yielding the equation of motion
\eqn\ncbsfteom{
\lform\,\varphi + 2\varphi\star\ln_\star\varphi = 0.
} 
In the context of BSFT, the spacetime noncommutativity of the tachyon
was derived in Refs.\refs{\Cornalba,\Okuyama} by taking into account
the effect of the constant $B$-field exactly in the worldsheet Green's
function of the fields $X^{1,2}$.

Since the commutative theory admits a gaussian soliton, it is natural
to try the ansatz \Gansatz\ for the solution of \ncbsfteom. However,
the second term poses some problem. Neither do we have an expression
for the star-deformed logarithm of an ordinary exponential function,
nor does the polynomial \apolynom\ seem to have a meaningful $p\to 1$
limit. Nevertheless, we can go back to the Eq.\ncpeom\ and take the 
$p\to 1$ limit on the ansatz \Gansatz. The first term is:
\eqn\boxphi{
\lim_{\epsilon\to 0}\left[{1\over\epsilon}(1+\epsilon)^{-\half\lform}
-{1\over\epsilon}\right]\,g(\hbox{\bf x})\; =\;
-{1\over2}\lform\,g(\hbox{\bf x})\; 
=\; 2a(1-ar^2)\,A^2e^{-ar^2}, }
the same as in the commutative case. Now, let us note the pattern of
terms in the $n$-fold star product of the gaussian in
\gaussstarn---the coefficients of $a\theta$ are determined by those 
in the expansion of $(1+x)^n$.  Ignorning the limit on the sum, as is 
appropriate for the non-integer power $1+\epsilon$, and dropping terms
of order $\epsilon^2$ or higher, we write:
\eqn\fracpower{
\left(\star g(\hbox{\bf x})\right)^{1+\epsilon} =
{A^{2(1+\epsilon)}\over 1 + \epsilon\left({(a\theta)^2\over 1.2} +
{(a\theta)^4\over 3.4} + \cdots\right)}\,\exp\left[-ar^2\left(
{1+\epsilon\left(1-{(a\theta)^2\over 2.3}-{(a\theta)^4\over 4.5} -
\cdots\right)\over
1+\epsilon\left({(a\theta)^2\over 1.2}+{(a\theta)^4\over 3.4} +
\cdots\right)}\right)\right].  }
{}From this we get the term in the RHS of the equation of motion:
\eqn\loggauss{
\eqalign{
\lim_{\epsilon\to 0}\left[{1\over\epsilon}\left(\star g(\hbox{\bf x})
\right)^{1+\epsilon}-{1\over\epsilon}\right] &=
\Bigg[\left(2\ln A - 
{(a\theta)^2\over1.2} - {(a\theta)^4\over3.4}-\cdots\right)\cr
&\qquad\; - 2ar^2\left(\half-{(a\theta)^2\over1.3}-
{(a\theta)^4\over3.5}-\cdots\right)\Bigg]\,
A^2e^{-ar^2}. }}
Comparing Eqs.\boxphi\ and \loggauss, we find that the amplitude is
determined by the transcentental equation:
\eqn\atrans{
2a = 1 - 2\left({(a\theta)^2\over1.3} + {(a\theta)^4\over3.5}
+ {(a\theta)^6\over 5.7} + \cdots\right)
= {1-(a\theta)^2\over 2a\theta}
\ln\left({1+a\theta\over 1-a\theta}\right), }
which interpolates between the width $a=1/2$ of the BSFT soliton and
that of the noncommutative GMS soliton $a=1/\theta$. 
The amplitude is determined in terms of $a$ as:
\eqn\lnA{
\eqalign{
2\ln A &= 2a + {(a\theta)^2\over1.2} + {(a\theta)^4\over3.4}
+ {(a\theta)^6\over5.6} + \cdots\cr
&= {1+a\theta\over 2a\theta}
\ln(1+a\theta) - {1-a\theta\over 2a\theta}\ln(1-a\theta). }}
This is also a smooth function of $\theta$ connecting the BSFT soliton
with $A=\sqrt{e}$ to the GMS one with $A=\sqrt{2}$.

The total energy of the gaussian lump $E=\half\int d^2x\,g\star g=
{\pi A^4/2a}$ is again a function of $\theta$ through Eqs.\atrans\
and \lnA, varying from $E_{\theta=0}=2\pi e^2$ to 
$E_{\theta\to\infty}=4\pi\theta$.
 
As an aside, let us attempt a naive power series expansion for
$g(\hbox{\bf x})\star\ln_\star g(\hbox{\bf x})$, which gives the series:
\eqn\ncrhs{
\eqalign{ 
& A^2\Bigg[ -\zeta(1)\exp(-a r^2) + \half{A^2\over(1+a^2\theta^2)}
\exp\left(-{2ar^2\over 1+a^2\theta^2}\right)\cr
&\qquad + {1\over 24}{A^4\over (1+a^2\theta^2)(1+3a^2\theta^2)}
\exp\left(-{3+a^2\theta^2\over 1 + 3a^2\theta^2}ar^2\right)
\cr
&\!\!\!\! 
- {1\over 72}{A^6\over (1+3a^2\theta^2)(1+6a^2\theta^2
+a^4\theta^4)}\exp\left(-{4+4a^2\theta^2\over 1 + 6a^2\theta^2
+a^4\theta^4}ar^2\right) +\cdots\Bigg].
}}
The above, in which we have used the $\zeta$-function regularization
for the infinite sums of powers of the natural numbers, should be
contrasted with the naive series expression of its commutative
analogue:
\eqn\crhs{
\eqalign{
A^2e^{-ar^2}\ln\left(A^2e^{-ar^2}\right) &= A^2\Bigg[
-\zeta(1)e^{-ar^2} + {1\over2}e^{-2ar^2}\cr
&\qquad\; + {1\over 24}e^{-3ar^2} - {1\over 72}e^{-4ar^2} + 
\cdots\Bigg].
}} 
For consistency, the series \ncrhs\ must sum to the RHS of
Eq.\loggauss. This identity gives the $\star$-logarithm of an ordinary
exponential function.


\newsec{Comments on multisolitons}
In addition to the gaussian \nclump, the Eq.\projector\ admits
other solutions which can be interpreted as multisolitons\GoMiSt.
However, these are unstable and reduce to \nclump\ at any finite
$\theta$. An interesting class of multisolitons were studied in 
\GoHeSp. Let
\eqn\twocentre{
f_{w_1w_2} = \exp\left(-{1\over\theta}(\bar z - \bar w_1)
(z - w_2)\right),
}
where $z=x^1+ix^2$ and we will refer to $(w_1,w_2)$ as the two
centres. The configuration
\eqn\GHSmulti{
\Pi=\sum_{i,j=1}^n A_{ij} f_{w_iw_j} = \sum_{i,j} A_{ij}
e^{-(\bar z - \bar w_i)(z - w_j)/\theta}
}
solves $\Pi\star\Pi=\Pi$ (for any value of $\theta$) for specific
choices of $A_{ij}$, and hence the equations of motion \projector\ of
a noncommutative scalar field theory at $\theta\to\infty$. By the
Moyal-Weyl correspondence $f_{w_iw_j}\sim |w_i\rangle\langle w_j|$,
where $|w\rangle \sim e^{w a^\dagger}|0\rangle$ is a {\it coherent
state}. The amplitudes of the multisoliton \GHSmulti\ are determined
in terms of the inner products of these states: $||A_{ij}|| \sim
||\langle w_i|w_j\rangle ||^{-1}$. To be more precise, Eq.\GHSmulti\
(given in terms of \twocentre), is the multisoliton for distinct
values of $w_i$ well-separated on the complex plane. When
$w_1,w_2,\cdots,w_\ell$ approach the same value, say, $w$, the
solution is in terms of projection operators in an $a$-invariant
subspace of the Hilbert space spanned by
$\left\{|w\rangle,a^\dagger|w\rangle,\cdots,
\left(a^\dagger\right)^{\ell-1}|w\rangle\right\}$. The multisolitons
of \refs{\GoHeSp} have a moduli space that survive to order 
${\cal O}(1/\theta)$. 

Inspired by this, let us define the more general two-centred gaussian
\eqn\twogauss{
f_{w_iw_j}(a_{ij},A_{ij})=A_{ij}\exp\left(-a_{ij}(\bar z - \bar w_i)
(z - w_j)\right),
}
where $a_{ij}$ and $A_{ij}$ are real. We will try a multisoliton
ansatz in terms of \twogauss\ for the equation of motion \ncpeom\ of
the $p$-tachyon. It is straightforward to check that:
\eqn\startwogauss{
\eqalign{
f_{w_1w_2}(a_{12},A_{12}) &\star f_{w_3w_4}(a_{34},A_{34})
= f_{{\tilde u}{\tilde v}}(\tilde a,\tilde A),\cr
\tilde{a} &= {a_{12}+a_{34}\over 1+\theta^2 a_{12}a_{34}},\cr
\tilde{A} &= {A_{12}A_{34}\over 1+\theta^2 a_{12}a_{34}}
\exp\left\{-{a_{12}a_{34}\over a_{12}+a_{34}}(\bar w_1-\bar w_3)
(w_2-w_4)\right\},\cr
{\tilde u} &=  {a_{12}(1+a_{34}\theta)\over 
a_{12}+a_{34}}\, w_1 + {a_{34}(1-a_{12}\theta)\over 
a_{12}+a_{34}}\, w_3,\cr
\tilde v &= {a_{12}(1-a_{34}\theta)\over a_{12}+a_{34}}\, w_2
+ {a_{34}(1+a_{12}\theta)\over a_{12}+a_{34}}\,w_4. 
}}
Thus, not only does the width and amplitude get modified as in 
\gaussstarn, the centres also get shifted. Thus, the $\star$-product
of $\sum_{ij} f_{w_iw_j}(a_{ij},A_{ij})$ with itself generates new
terms. 

The LHS of \ncpeom, on the other hand, modifies only the width and
amplitude\foot{One may say the differential operator maps operators
of the form \twogauss\ to a different quantum mechanical Hilbert space
with a modified value of the Planck constant.}:
\eqn\lhseom{
e^{-{1\over2}\ln p\,\lform}f_{w_iw_j}(a_{ij},A_{ij})
= f_{w_iw_j}\left({a_{ij}\over 1+2a_{ij}\ln p},{A_{ij}\over 
1+2a_{ij}\ln p}\right),}
but leaves the centres unchanged. Thus the multisolitons of the type
found in Ref.\GoHeSp\ do not solve the equation of motion
\ncpeom\ of the $p$-tachyon. However, since the $p\to 1$ limit is the
usual bosonic strings, one would expect to have multisolitons in this
limit. It would be interesting to see how that happens.

The centres do not shift in the Eq.\startwogauss\ if
$a_{ij}=1/\theta$, or if $w_1=w_3$ and $w_2=w_4$. In both cases,
$\tilde u= w_1$ and $\tilde v= w_4$ and hence $f_{w_1w_2}\star
f_{w_3w_4}\sim f_{w_1w_4}$.  The former case is what we expect from
the Moyal-Weyl correspondence. In particular, this is true in the
$\theta\to\infty$ limit recovering the multisoliton of \GoHeSp. In the
second case, we find that $f_{ww'}(a,A)$ solves the equation of motion
\ncpeom\ of the $p$-tachyon with $a$ and $A$ given by \apolynom\ and
\amplitude\ respectively. However, this solution is complex, unless
$w=w'$, in which case we find a simple generalization of the gaussian
solution of Sec.~3 centred around the point $w$ in the $z$-plane. This
corresponds to the moduli due to the translation zero-modes of the 
solitons (found in Refs.\refs{\FrNi,\GhSeP} for \plump\ at $\theta=0$). 


\newsec{Concluding remarks}
We obtained gaussian lump solutions to the equations of motion of a
noncommutative deformation of the tachyon of the $p$-adic string
theory for all values of the noncommutative parameter $\theta$.  It is
shown that this one parameter family of solutions survives the $p\to
1$ limit, in which the $p$-adic string approximates the BSFT
description of the usual bosonic string. The solutions interpolate
smoothly between the soliton of the commutative $p$-adic tachyon
(respectively BSFT) and the GMS soliton in the limit of infinite
noncommutativity. This is unlike other exact solutions obtained in
field theories of scalars and gauge fields, see {\it e.g.} \NCSoln,
which do not have a smooth commutative limit.

It should be mentioned that the similarity between the solitons of the
$p$-string theory and the noncommmutative field theories have been
noticed in Ref.\DrVo. In particular, it was pointed out that the
expression \nclump\ for $\theta=4\ln 2$ is identical to solitonic
2-brane of the 2-adic string theory. However, this seems to be a
fortuitous coincidence without any particular significance. Indeed,
the GMS soliton \nclump\ is a solution near $\theta\to\infty$.

We have introduced noncommutativity in the $p$-adic string theory by
replacing ordinary products with the Moyal star product in the
spacetime effective action of its tachyon. This is admittedly rather
ad hoc.  In usual string theory this deformation is due to a constant
background of the second rank antisymmetric tensor field $B$, which
alters the worldsheet properties of the fields $X^{1,2}$. This can be
taken into account exactly in the (worldsheet) Green's functions of
these fields. The `worldsheet' of the $p$-adic string is an infinite
Bethe lattice\Zabro. It would be interesting to see if the lattice
worldsheet action admits a deformation by the $B$-field that induces
the spacetime noncommutativity. We hope to report on this in
future\refs{\GhKa}.

Ref.\Schom\ noted the similarity between the spacetime
noncommutativity arising out of a $B$-field and the methods of
deformation quantization.  It would nice if noncommutative $p$-tachyon
can be used in the quantization of its nonlocal field theory.

Finally, the $p$-adic string seems to have many of the features in
common with the usual string theory. Moreover, there are direct
connections through the adelic relations\FrWi\ and the $p\to 1$
limit\GeSh.  For these reasons, it may be worthwhile to develop a
better understanding of the $p$-adic string theory. As of now, we only
know some properties of its D-branes in flat space. Even there the
tension of the D-brane has not been computed from a worldsheet
description. This, as well as other problems, {\it e.g.}, the
$B$-field background that we studied in this paper, call for a proper
understanding of the closed $p$-adic string.

%
%
\bigskip

\noindent{\bf Acknowledgement:} We are grateful to Suresh Govindarajan,
Mathew Headrick, Teruhiko Kawano, Ashoke Sen and especially Rajesh
Gopakumar for useful discussion.  We would like to thank Tohru Eguchi
and the Japan Society for the Promotion of Science (JSPS) for an
Invitation Fellowship which partially supports this work.




\listrefs 

\bye